\documentclass[10pt,prd,longbibliography,superscriptaddress]{revtex4-2}
\usepackage{graphicx}
\usepackage{epstopdf,subfigure}

\newtheorem{thm}{Theorem}

\newcommand{\cn}{\mathop{\rm cn}\nolimits}

\newcommand{\sech}{\mathop{\rm sech}\nolimits}
\newcommand{\csch}{\mathop{\rm csch}\nolimits}
\def\di{\displaystyle}

\begin{document}


\markboth{Agaoglou, Rothos, Frantzeskakis, Veldes \& Susanto}{Bifurcation results for travelling waves in nonlinear magnetic metamaterials}

\title{Bifurcation results for travelling waves in nonlinear magnetic metamaterials}

\author{M.\ Agaoglou}
\address{Department of Mathematics, Physics Computational Sciences, Faculty of Engineering,\\ Aristotle University of Thessaloniki, Thessaloniki 54124, Greece}

\author{V.M.\ Rothos}
\address{Department of Mathematics, Physics Computational Sciences, Faculty of Engineering,\\ Aristotle University of Thessaloniki, Thessaloniki 54124, Greece}

\author{D.J.\ Frantzeskakis}
\address{Department of Physics, University of Athens, Panepistimiopolis, Zografos, Athens 15784, Greece}

\author{G.P.\ Veldes}
\address{Department of Physics, University of Athens, Panepistimiopolis, Zografos, Athens 15784, Greece}

\author{H.\ Susanto}
\address{Department of Mathematical Sciences, University of Essex, Wivenhoe Park, Colchester CO4 3SQ,\\ United Kingdom}


\begin{abstract}
In this work we study a model of a one-dimensional magnetic metamaterial formed by a discrete array of
nonlinear resonators. We focus on periodic and localized travelling waves of the model, in the presence of
loss and an external drive. Employing a Melnikov analysis we study the existence and persistence of
such travelling waves, and study their linear stability. We show that, under certain conditions, the presence
of dissipation and/or driving may stabilize or destabilize the solutions.
Our analytical results are found to be in good agreement with direct numerical computations.
\end{abstract}
\maketitle

\section{Introduction}

Magnetic metamaterials (MMs) are metamaterials consisting of periodic arrays of split-ring resonators
(SRRs), in one-, two- and three dimensions \cite{review1}. Both the dimensions of SRRs and their
inter-space distance are small relative to the free space wavelength and, thus, the quasi
magnetostatic approximation \cite{review2} governs the dynamics of electromagnetic (EM) fields in
such settings. As shown in earlier works \cite{Shamon1, Shamon2}, MMs support magneto-inductive (MI)
waves, due to the coupling between SRRs (which can be modelled as LC circuits). In fact, MI waves are
found in the short wavelength limit, which means that MI waves are slow waves: their
phase velocity and the group velocity are smaller than the velocity of light in free space. Magnetic
metamaterials have attracted much interest as they have already been used for the realization of
various microwave devices, including couplers and splitters \cite{Shamon3}, shifters \cite{Nefedov},
delay lines \cite{Freire1}, parametric amplifiers \cite{Sydoruk}, magneto-inductive lenses
\cite{Freire2}, polarizers \cite{Gansel}, and so on.

Nonlinear MMs, which can be implemented by embedding the array of SRRs into a nonlinear dielectric
\cite{Zharov,Shadrivov} or by inserting diodes into resonant conductive elements
\cite{Lapine,IVShadrivov}, have also been studied in various works.
The latter method can more easily be implemented in practice --see, e.g., relevant experimental works
on three-dimensional (``metacrystal'') MMs \cite{Huang} and on the modulation instability in MM
waveguides \cite{Tamayama}. In such settings, various studies have been performed, including the
localization of EM energy and the formation of discrete breathers \cite{Lazarides}, magneto-inductive
envelope solitons \cite{Kourakis} and gap solitons \cite{Cui}, transfer of EM power in MM
transmission lines \cite{Lazarides2} and others.

In this paper, we consider periodic and localized travelling waves in a one-dimensional (1D)
nonlinear MM, composed by a chain of nonlinear SRRs. In particular, we study the
persistence of the waves in the presence of perturbations, namely dissipations and travelling drives.
This is done by employing a Melnikov-type of analysis, which is complemented by a stability analysis of the solutions. The present study extends our previous work \cite{1} on the analysis of a 1D MM with Kerr-type materials by considering a quadratic nonlinearity of the system. More specifically, our presentation is organized as follows. In Section 2, we present the physical model.
In Section 3, we study periodic travelling waves under the assumption that the unperturbed system has periodic solutions.
We find conditions under which these periodic solutions persist using the subharmonic Melnikov bifurcation method.
In Section 4, we consider the persistence of a localized wave that is the limiting case of the periodic solutions.
In Section 5, we present numerical computations comparing the analytical results in the preceding sections.
Additionally, we also consider the existence and stability of solutions due to the periodic travelling drive,
whose amplitude is limited by the magnitude of the forcing. Finally, Section 6 summarizes our findings.

\section{The nonlinear magnetic metamaterial model}
\begin{figure}[htbp]
\centering
\includegraphics[width=10cm]{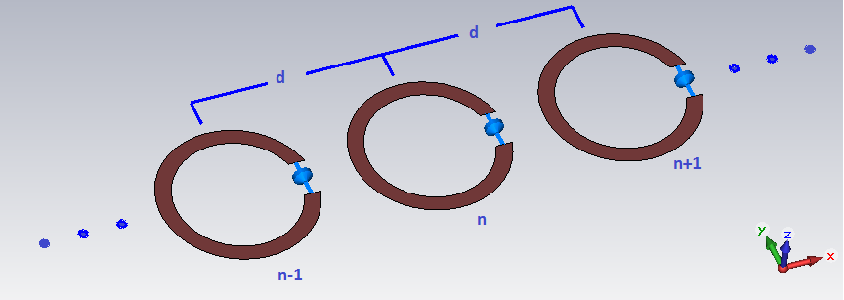}
\includegraphics[width=10cm]{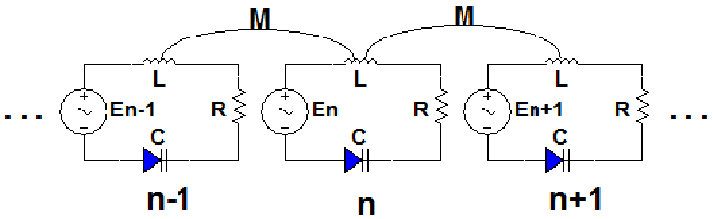}
\caption{Top panel: A representative illustration of the nonlinear magnetic metamaterial structure.
Bottom panel: the equivalent circuit model (see text for details).}
\label{fig:model}
\end{figure}
We consider a nonlinear MM, composed by an infinite 1D chain of identical nonlinear circular SRRs,
as shown in the top panel of Fig.~\ref{fig:model}. Each of the SRRs is modelled by a resonant
circuit, namely a LC circuit, and the nonlinearity in the model is assumed to be introduced by
inserting a diode in the slit of SRR --cf. bottom panel of Fig.~\ref{fig:model}; this way, the
total effective capacitance of SRR becomes nonlinear (see details below). Furthermore, each SRR in
the chain is driven by the electromotive force $E_n(t)$, which is induced by an external EM field.
The SRRs are assumed to be magnetically coupled, i.e., via the mutual inductance $M$ between two
inductors; here, solely next-nearer-neighbour such interactions
will be considered. Finally, we note that dissipative effects will also be taken into account,
accounted for by the finite conductivity of the metallic SRR structure.

Let us now consider the electric analogue of the model. According to, e.g. Ref.~\cite{Lapine}, each
SRR can be modelled by an effective RLC circuit, characterized by the ohmic resistance $R$, the self-
inductance $L$, and a capacitance $C$, which is a combination of the linear capacitance $C_{\rm sl}$
due to the slit of the SRR and the effective capacitance of the inserted diode \cite{Carbonell}.
The latter is biased at a constant voltage, say $U_0$, and its capacitance is nonlinear,
depending on the voltage $U$ applied across it. Assuming that this voltage
$U$ does not change significantly from the bias voltage $U_0$, we can Taylor expand the effective
capacitance $C(U_n)$ of the $n$-th SRR taking into regard only the lowest order terms, namely,
\begin{equation}
C(U_{n}) \approx C_{0}+C_{0}^{'}U_n=C_0(1+\alpha U_n),
\label{eq:cofv}
\end{equation}
where $C_0 \equiv C(U_{0}) + C_{\rm sl}$, $C_{0}^{'}\equiv(dC/dU)\mid_{U_{0}}$,
and $\alpha=C_0^{'}/C_0$. Taking into regard the above expression, the application of
Kirchhoff's voltage law for the $n$-th SRR (see bottom panel Fig.~\ref{fig:model}) leads to the
following equation for the voltage:
\begin{eqnarray}
U_n+L\frac{dI_{n}}{dt}+RI_n-M\frac{dI_{n-1}}{dt}-M\frac{dI_{n+1}}{dt}-E_n(t)=0,
\label{eq:KVL}
\end{eqnarray}
where
\begin{eqnarray}
I_n&=&\frac{dQ_n}{dt}=\frac{d}{dt}[C(U_n)U_n],
\label{eq:Q1}
\end{eqnarray}
is the current of the $n$-th SRR, and we have taken into regard that the mutual inductance $M$
is negative in the coplanar configuration of Fig.~\ref{fig:model}.

For our considerations below, we will now derive an equation for the charge $Q_n$ of the capacitor of
the $n$-th SRR, similarly to what was done in Ref.~\cite{Lazarides2}. Taking into regard that the
charge $Q_n$ is given by:
\begin{eqnarray}
Q_{n}&=&C{(U_n)}U_n=C_0(1+\alpha U_n)U_n,
\label{eq:charge}
\end{eqnarray}
we solve Eq.~(\ref{eq:charge}) with respect to $U_n$ up to the order $\mathcal{O}(Q_n^2)$, and
substituting the resulting expression into Eq.~(\ref{eq:KVL}), we obtain the following
equation for $Q_n$:
\begin{equation}
L\frac{d^2}{dt^2}\left(Q_n-\lambda Q_{n-1}-\lambda Q_{n+1}\right)+R\frac{dQ_n}{dt}+U_n-E_n(t)=0,
\label{eq:model_Q01}
\end{equation}
where $\lambda=M/L$ is the coupling coefficient. Next, we introduce the scale transformations
$u_n= U_n(U_c)^{-1}$, $q_n= Q_n(C_0 U_c)^{-1}$ (where $U_c$ is the breakdown potential of the diode), 
%
%
%
%
and $t \rightarrow \omega_{0} t$ [where $\omega_0^{2}= (LC_{0})^{-1}$], and cast Eq.~(\ref{eq:model_Q01}) into the following form:
\begin{equation}
\frac{d^2}{dt^2}\left(q_n-\lambda q_{n-1}-\lambda q_{n+1}\right)+\gamma \frac{dq_n}{dt}+q_n
- \beta q_n^2-\varepsilon_n(t)=0,
\label{eq:modelQn}
\end{equation}
where $\beta = \alpha U_c$, $\gamma=RC_0\omega_0$ is the loss coefficient and $\varepsilon_n= U_c^{-1}E_n$ is the normalized electromotive force.

\section{Bifurcation results for Travelling Waves}
%

In this section, we focus on
a travelling drive $h$, which is $C^{2}-$smooth and $2\pi-$periodic, i.e.,
\begin{equation}
\varepsilon_n(t)= h(z), \quad z=\omega t+pn,
\label{td}
\end{equation}
where $\omega >0$ is the external driving frequency, $p\neq 0$ is the wavenumber of the driving wave field, $n\in Z$ and $t\in R$ is the evolution time.
Assuming that $q_{n} (t)=U(z)$, with $U\in C^{2} (R ,R)$, we obtain from
(\ref{eq:modelQn})
the following advance-delay equation:
%
\begin{eqnarray}
\label{3}
&&\omega^{2} U^{''}(z)+U(z)-\beta U^2(z) \nonumber \\
&&-\epsilon \lambda \omega^{2} [U^{''} (z-p) +U^{''}
(z+p)] 
+\epsilon\gamma\omega U^{'}(z)-\epsilon h(z)=0,
\end{eqnarray}
where we have considered that parameters $\gamma$, $\lambda$ and the drive 
are small, $\propto \epsilon$ (where $\epsilon$ is a nonzero formal small parameter).

\subsection{Case 1: periodic travelling waves}
We now make the following assumptions:
%
\begin{itemize}
\item[(H1)]
$U^{''} (z)-\varphi(U(z))=0$ (where $\varphi$ is the nonlinearity of the magnetic material ,which is normally assumed to be of Kerr-type \cite{Lazarides} and in our case $\varphi =-U(z)+\beta U^2(z) $) has a $\overline{T}-$ periodic solution $U_{0}$.
\end{itemize}

\textit{Remark 3.1.} Since $U_{0} (-z+c_{0})$ also solves $U^{''} (z) +U(z)-\beta U^2(z)=0$
and there is $z_{0}\in R$ such that $U^{'}_{0} (z_{0}) =0$,
we may suppose that $U_{0} (0) =0$ and then $U_{0} (z) =U_{0} (-z)$. Then
\[U_{\omega} (z):= U_{0} (z/\omega)\]
satisfies $\omega ^{2} U^{''}_{\omega} (z) +U_{\omega} (z)-\beta U_{\omega}^2(z)=0$. Note $U_{\omega}$ is $T_{\omega} :=\overline{T} \omega -$periodic and even.

We also assume the resonance condition
\begin{itemize}
\item[(H2)] $T_{\omega}=2\pi{u}/{v}$ for $u,v\in N$, and because of that
we will study the case with $v=1$.
\end{itemize}
The assumption (H2) then becomes
\begin{itemize}
\item[(H3)] $\displaystyle\omega ={2\pi u}/{\overline{T}}$ for $u\in N$.
\end{itemize}

We now
apply the standard subharmonic Melnikov method to (\ref{3}) based on the Lyapunov-Schmidt method \cite{1}; we compute that the Melnikov function is given by:
\begin{equation}\label{meln}
M^{u} (a)=-\gamma \int_{0}^{\overline{T}} U^{'}_{0}(z)^{2}\,dz -\frac{2\pi u}{\overline{T}}\int_{0}^{\overline{T}} h^{'} \left(\frac{2\pi u}{\overline{T}} z
+a\right)U_{0} (z)\,dz.
\end{equation}
From (H1), we obtain
$$\frac{1}{2} (U^{'})^{2} +\left(\frac{U^{2}}{2} -\frac{\beta U^{3}}{3}\right)=c_{0}.$$
where $c_{0}$ is a constant.
The above equation has a one-parameter family of periodic solutions \cite{2}
$$\di{U_{0,\alpha} (z)=A\frac{1-{\rm cn}(\frac{z}{g})}{1+{\rm cn}(\frac{z}{g})} +\alpha,}$$
where ${\rm cn}$ is the cnoidal Jacobi elliptic function, and $g$, $\alpha$ are
constants (see details in Appendix A).

For $\alpha >0$ with periods $\overline{T}=\overline{T(\alpha)}=4K(k)g$
(where $K(k)$ is the complete elliptic function of the first kind, and $k$ the elliptic modulus),
we can compute the first term on the right hand side of the Melnikov function (\ref{meln}) as
$$\int_0^{\overline{T}(\alpha)} U^{'}_{0,\alpha}(z)^{2}\,dz=\frac{4A^{2}}{g^{2}}
\int_0^{\overline{T}(\alpha)} \frac{ {\rm sn}^{2}(\frac{z}{g}){\rm dn}^{2} (\frac{z}{g})}
{(1+{\rm cn}(\frac{z}{g}))^{4}} \,dz=\frac{4A^{2}}{g} \int_0^{4K(k)} \frac{{\rm sn}^{2}(z)
{\rm dn}^{2} (z)}{(1+{\rm cn} (z))^{4}} \,dz$$
$$=\frac{4A^{2}}{g} \frac{((-1 + 2 k) E(
    {\rm am}(4 K(k), k), k) - 4 (-1 + k) K(k))}{3 (1 +
   {\rm cn}(z))^{4} k}.$$
where ${\rm sn}$ and ${\rm dn}$ are Jacobi elliptic functions and ${\rm am}$ is the Jacobi amplitude.\\
Similarly, by taking
\begin{equation}
h(z)=\cos z,
\label{hz}
\end{equation}
we can evaluate the remaining term of the function (\ref{meln}) as explained in Appendix B; the result is:
\begin{eqnarray}
\frac{2\pi u}{\overline{T}(\alpha)}\int_0^{\overline{T}(\alpha)} h^{'}\left(\frac{2\pi u}{\overline{T(\alpha)}}z+ a\right)U_{0,\alpha} (z)\,dz =\frac{A\pi u}{2K(k)}\Xi \sin a , 
\end{eqnarray}
%
%
%
%
%
%
%
%
where
\begin{equation}
\Xi=\frac{P}{24 \pi^{5} u^{5} K^{5}(k)},
\end{equation}
and
\begin{eqnarray}
P &=& -(64 \pi u \cos(2 \pi u K^2(k))K(k)^{2} (24 - 48 k + \pi^{2} u^{2} K^{2}(k) \nonumber \\
&\times& (-3 + 16 (-1 + 2 k) K(k)^{2})) 
+32 (-24 + 48 k + \pi^{2} u^{2} K^{2}(k) \nonumber \\
&\times&(3 + K^{2}(k) (48 - 96 k + 2 \pi^{2} u^{2} K^{2}(k) (-3 + 8 (-1 + 2 k) K^{2}(k)))))
\nonumber \\
&\times& \sin(2 \pi u K(k)^{2}))}{.
\end{eqnarray}
%
%
%
Here, we have used the fact that $\cos (z), cn(z)$  are even and $\sin (z)$ is odd and
made the expansion of $\di{-1+\frac{2}{(1+cn(z))}}$.

Summing up the contributions above, the Melnikov function is then given by
\begin{eqnarray}
M^{u}(a)=& -&\gamma \int_{0}^{\overline{T}} U^{'} _{0} (z)^{2}\,dz -\frac{2\pi u}{\overline{T}}
\int_{0}^{\overline{T}} h^{'} \left(\frac{2\pi u}{\overline{T}} z+a\right)U_{0} (z)\,dz =\nonumber \\
&-&\gamma\frac{4A^{2}}{g} \left[\frac{((-1 + 2 k) E(
    {\rm am}(4 K(k), k), k) - 4 (-1 + k) K(k))}{3 (1 +
   {\rm cn}(z))^{4} k}\right]\nonumber \\
   &+&\frac{A\pi u}{2K(k)} \Xi\sin a .
\end{eqnarray}

\begin{thm}
Suppose $(H1)$ and $(H2)$. If there is a simple zero $a_{0}$ of a Melnikov function,
i.e $M^{u/v}$ =0 and $D_{a} M^{u/v} (a_{0})\neq 0 $, then there is a $\delta >0$ such that for
any $0\neq \varepsilon \in (-\delta,\delta)$ there is a unique $2\pi u-$periodic solution $U(z)$ of (\ref{3}) with
$$U(z)= U_{0} (\frac{z-a_{0}}{\omega}) +O(\varepsilon)$$
\end{thm}

To have a simple zero of $M^{u} (a)$, we need
$$\gamma \Lambda (\alpha, u)<1,$$
where
\begin{large}
   $$\Lambda (\alpha, u):=\frac{\frac{4A^{2}}{g} [\frac{((-1 + 2 k) E(
    am(4 K(k), k), k) - 4 (-1 + k) K(k))}{3 (1 +
   cn)^{4} k}]}{\frac{A\pi u}{2K(k)} \Xi}$$
\end{large}
%
%
%
Thus, we see that
\begin{equation}
\gamma<\frac{1}{\Lambda (\alpha,u)}
\label{rm1}
\end{equation}
gives the magnitude for the damping in order to apply Theorem 1.

\subsection{Case 2: localized travelling waves}

For localized travelling waves, instead of (H1) we have the following assumption:
\begin{itemize}
\item[(C1)] $\varphi (0)=0, \varphi ^{'} (0)<0$ and $U^{''} (z) -\varphi (U(z)) =0$
(where in our case $\varphi=-U(z) +\beta U^{2}(z)$) has an asymptotic localized solution
$\Gamma \in C^{2} (R,R)$ such that $\lim_{\mid z\mid \to +\infty} \Gamma (z)=0$
and $\lim_{\mid z\mid \to +\infty} \Gamma ^{'} (z)=0$.
\end{itemize}

Now, we apply the standard homoclinic Melnikov method to Eq.~(\ref{3}),
based on the Lyapunov-Schmidt method \cite{1}, and we get the Melnikov function:
\begin{equation}\label{6}
M(a)=\int_{-\infty}^{\infty} (-\gamma \Gamma^{'} (z) +h(\omega z+a))\Gamma ^{'} (z)\,dz.
\end{equation}


The unperturbed equation possesses a homoclicic solution:
$$\Gamma (z)=\frac{1}{\beta} -\frac{3\sech ^{2} (\frac{z}{2})}{2\beta}.$$

Using again Eq.~(\ref{hz}) for the form of the drive, the Melinkov function (\ref{6}) becomes:
\begin{eqnarray}
M(a)&=&\int_{-\infty}^{\infty} (-\gamma \Gamma^{'} (z) + h(\omega z+a))\Gamma^{'}(z)\,dz
\nonumber \\
&=&-\frac{6(\gamma +5\beta \pi \omega ^{2} \csch (\pi \omega) \sin (a))}{5\beta ^{2}}.
\end{eqnarray}
\begin{thm}
Suppose $(C1)$. If there is a simple zero $a_{0}$ of the Melnikov function
$$M(a)=\int_{-\infty}^{\infty} (-\gamma \Gamma^{'} (z) +h(\omega z+a))\Gamma ^{'} (z)\,dz$$
then there is $\theta>0$ such that for any $0\neq \varepsilon \in (-\theta,\theta)$ there
is a unique bounded solution $U(z)$ of (\ref{3}) on $R$ with
$$U(z)= \Gamma (\frac{z-a_{0}}{\omega}) +O(\varepsilon)$$
\end{thm}
Therefore, if we want the Melnikov function $M(a)$ to have a simple zero we need
\begin{equation}
\gamma< 5\beta \pi \omega ^{2} \csch(\pi \omega).
\label{rm2}
\end{equation}

Note that $M(a)$ gives a kind of $O(\epsilon)$-measure of the distance
between the stable and unstable manifolds of the periodic
solution $U_{0}$ of our perturbed system
which is at a $O(\epsilon)$-distance from $U_{0}$. Thus, if $ M(a)$ has a simple zero
at some points, as in our case, the implicit function theorem implies that these two manifolds
intersect transversally along a solution $\varphi$ which is homoclinic to
$U_{0}$.

\section{Numerical results}

To illustrate the theoretical results obtained in the previous sections, we have solved the
model equation in a moving coordinate frame [cf. Eq.~(\ref{3})] numerically.
The advance-delay equation under consideration is solved using a pseudo-spectral method, i.e., we express the solution $U$ in a Fourier series
\begin{equation}
U(z)=\sum_{j=1}^{J}\left[A_j\cos\left((j-1)\tilde{k}z\right)+B_j\sin\left(j\tilde{k}z\right)\right],
\label{ser}
\end{equation}
where $\tilde{k}=2\pi/L$ and $-L/2<z<L/2$. The Fourier coefficients $A_j$ and $B_j$
are then found by requiring the series to satisfy (\ref{3}) at several collocation points. Hence,
$2J$ collocation points are required, which are chosen with uniform grid points. Typically, we use
$J=50$. It is important to note that the physically relevant range for the coupling parameter
$\lambda$ is $|\lambda|<1/2$ \cite{1}.

The stability of a solution obtained from (\ref{3}) is then determined numerically through calculating its Floquet multipliers $\chi$, which are eigenvalues of the monodromy matrix. As a first-order system, the linearized equation of (\ref{eq:modelQn}) is:
\begin{equation}
\begin{array}{lll}
&\displaystyle\dot{u}_n= v_n, \\
&\displaystyle\dot{v}_n -\lambda \dot{v}_{n-1} -\lambda \dot{v}_{n+1} = -\gamma v_n -u_{n} +2\beta U_n u_{n},
\end{array}
\label{lin}
\end{equation}
where $U_n(t)=U(\omega t+np)$ (cf.\ (\ref{ser})) is a periodic solution of (\ref{eq:modelQn}). The linear system is integrated using a Runge-Kutta method of order four with periodic boundary conditions over the period $\overline{T}=L/\omega$. The monodromy matrix $\mathcal{M}$ is defined as \[
\left(
\begin{array}{ccc}
\{u_n(T)\}\\
\{v_n(T)\}
\end{array}
\right)=
\mathcal{M}
\left(
\begin{array}{ccc}
\{u_n(0)\}\\
\{v_n(0)\}
\end{array}
\right).
\]
A periodic solution is unstable if there is any Floquet multiplier that is
strictly greater than one in modulus. The typical dynamics of the instability is demonstrated by integrating the governing equation (\ref{eq:modelQn}) using the same numerical method.

Without loss of generality, below we fix
$\beta=1$. Considering a simple periodic drive in (\ref{hz}), we take
\begin{equation}
h(z)=\Delta\cos(z),
\end{equation}
where $\Delta\in R$ is the driving amplitude.

\subsection{Persistence and stability of traveling waves of the unperturbed system}

\begin{figure}[tbhp!]
\begin{center}
\subfigure[\,$\Delta=\gamma=\lambda=0$]{\includegraphics[width=0.45\textwidth]{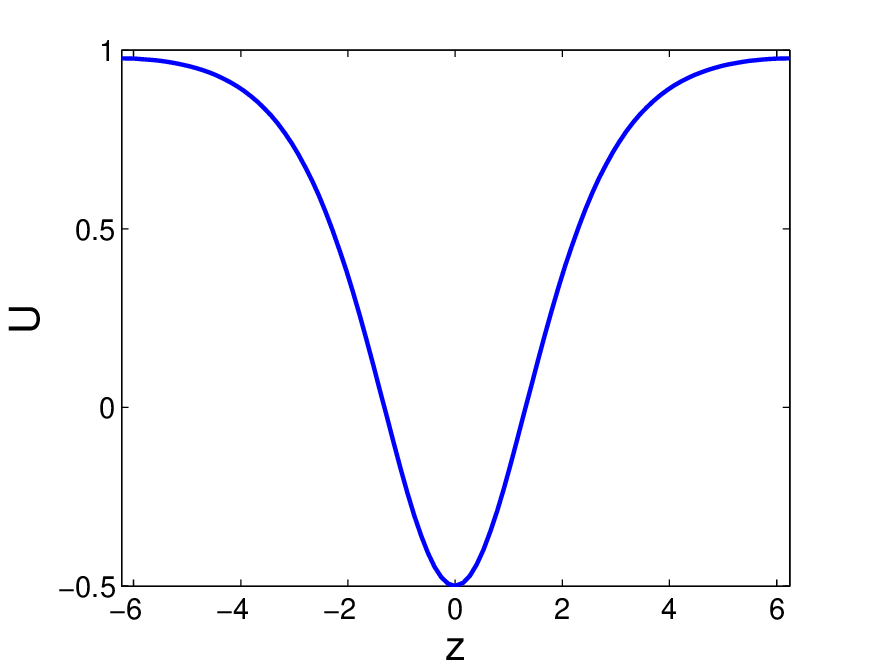}}\\
\subfigure[\,$\gamma=\lambda=0$]{\includegraphics[width=0.45\textwidth]{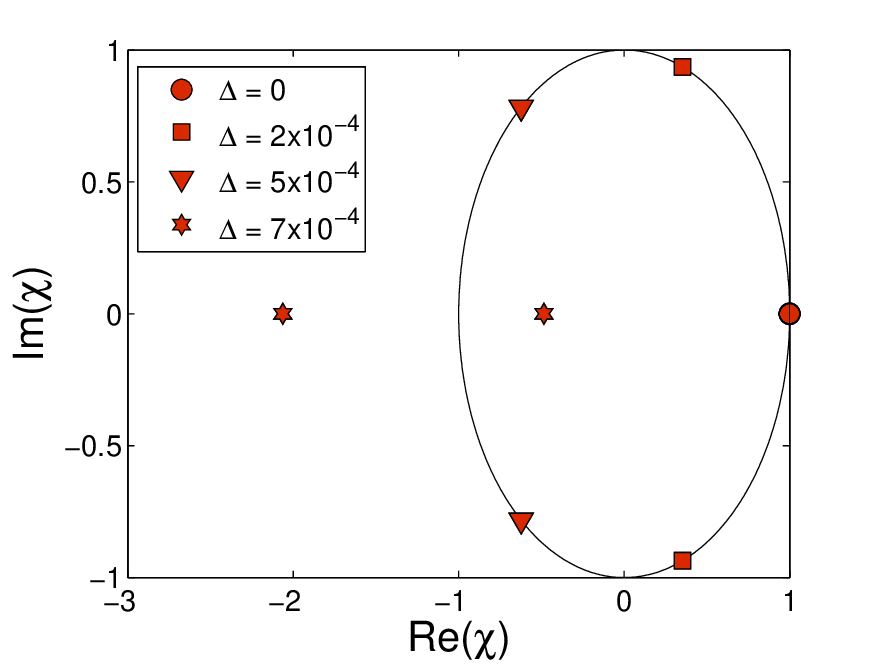}}\\
\subfigure[\,$\gamma=0$]{\includegraphics[width=0.45\textwidth]{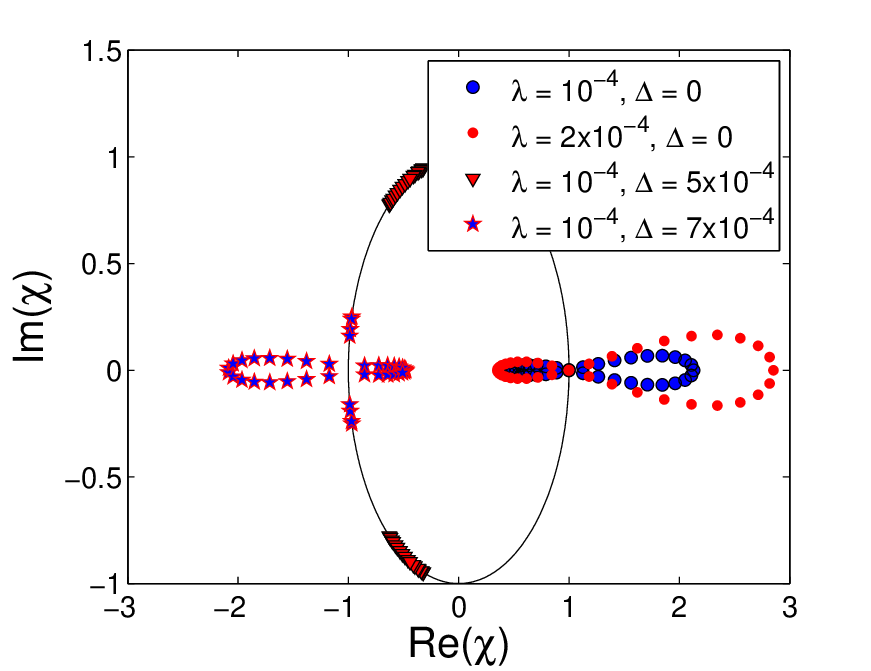}}
\end{center}
\caption{(a) A periodic wave in a traveling frame. (b)-(c) Floquet multipliers of the solution corresponding to panel (a) for parameter values as given in the captions. The solution period in all the plots is $\overline{T}=4\pi$. As a guide we also plot the unit circle in thin solid curve. The stability is calculated using 20 sites.}\label{fig1}
\end{figure}

In this part, we will study the persistence of periodic or localized travelling waves of the unperturbed system in the presence of small perturbations described in the preceding sections and the stability of the perturbed solutions. To do so, it is firstly important to understand the effect of each perturbation term in the governing equation (\ref{eq:modelQn}). 

In Fig.~\ref{fig1}(a) we show the solution profile of the advance-delay equation (\ref{3}) with
$\Delta=\gamma=\lambda=0$ and $\overline{T}=4\pi$. Turning on the perturbation by setting
only one of the parameters ($\Delta$, $\gamma$, or $\lambda$) nonzero, one would expect that the
solution will persist. This is indeed the case, except from the case of
$\gamma\neq0,\,\Delta=0$; physically speaking, this can be
understood by the fact that
damping will dissipate energy and drive is needed to
compensate the dissipation. Mathematically this is in agreement with Theorem~\ref{rm1},
which will be discussed further.

When the perturbation is small enough, the profile of the solution will be very similar to the
unperturbed case. However, the stability can change drastically. Shown in Fig.\ \ref{fig1}(b)
are the Floquet multipliers of the perturbed solution when the oscillators are decoupled, i.e.\
$\lambda=0$. Using several values of $\Delta$ (and $\gamma=0$),
we conclude that the drive pulls the multipliers along the unit circle and later destabilize the
solution through a period-doubling bifurcation. As for the effect of coupling on the stability of
solutions, we plot in Fig.~\ref{fig1}(c) the Floquet multipliers of the solution corresponding to
that in panel (a) for two values of $\lambda$ and $\gamma=0$. As seen, the coupling creates a band of
multipliers and for the present parameter values destabilizes solutions. Combining the two effects,
the dynamics of the multipliers are nontrivial and one may obtain either a stable or unstable
periodic solution, as depicted also in panel (c). When we obtain stable solutions, the stabilization
is clearly due to the drive. This result is similar to the stabilization of unstable discrete
solitary waves in parametrically driven oscillators reported in Refs.~\cite{syaf10,susa06}.

As for the dissipation, it may yield asymptotic stability. However, to stabilize unstable solutions
by increasing the dissipation parameter, the solutions may already cease to exist before all
corresponding unstable Floquet multipliers enter the unit circle. This is due to the bound that is
predicted by (\ref{rm1}); see also (\ref{rm2}) for localized waves. In the following, we will
compare the prediction with our numerics.

\begin{figure}[tbp]
\begin{center}
\subfigure[]{\includegraphics[width=0.45\textwidth]{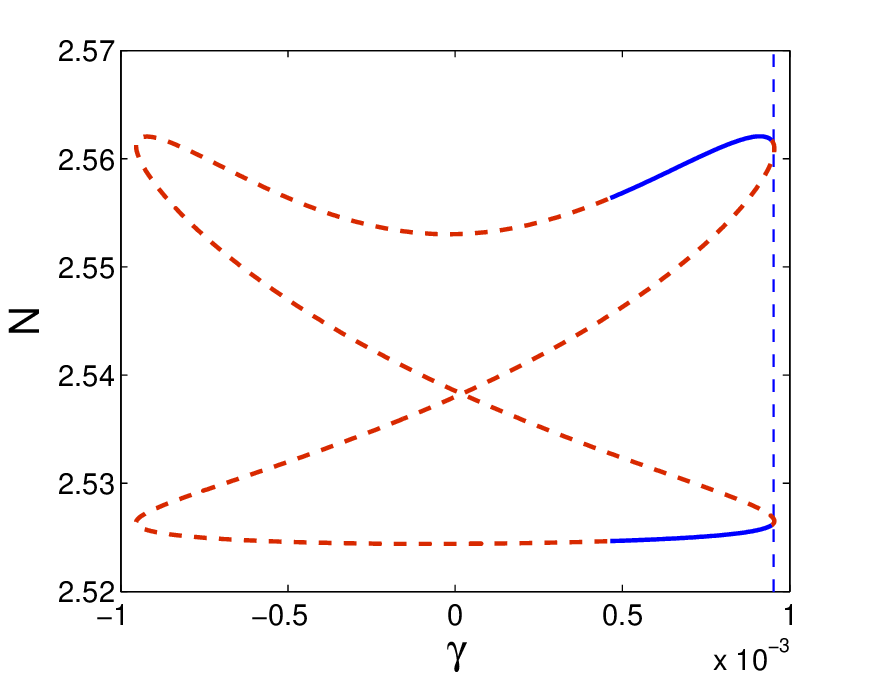}}\\
\subfigure[]{\includegraphics[width=0.45\textwidth]{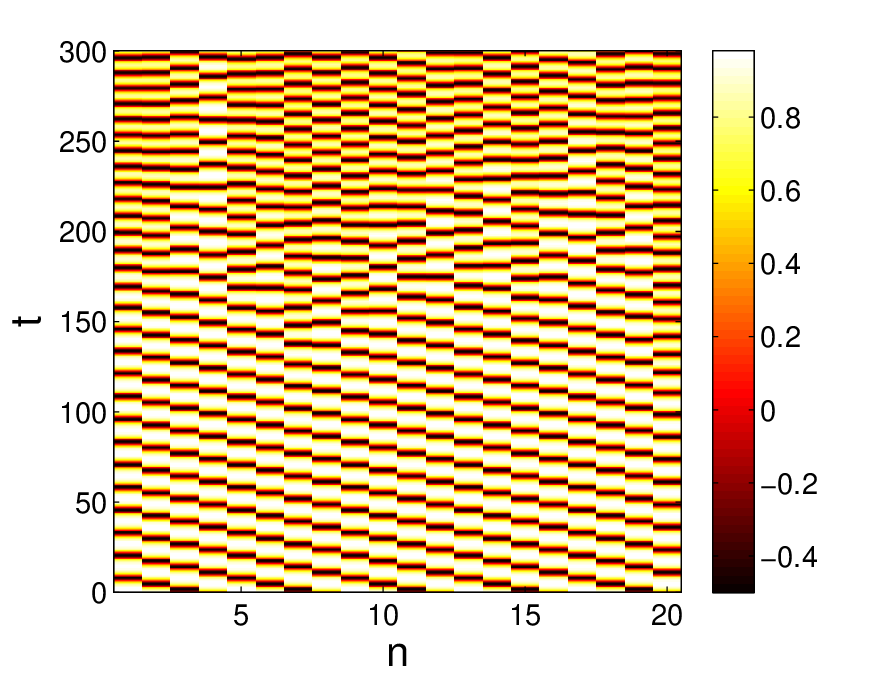}}
\end{center}
\caption{(a) Bifurcation diagram of the solution corresponding to Fig.\ \ref{fig1}(a) with $\Delta=7\times10^{-4}$ and $\lambda=10^{-4}$. The vertical axis is the solution norm $N=\sqrt{\int_0^{\overline{T}}U(z)^2\,dz}$. Solid and dashed curve indicate unstable and stable solutions, respectively. The stability is calculated using $20$ sites. The vertical dashed line is the prediction (\ref{rm2}). (b) Typical dynamics of the instability. The initial condition is the solution obtained immediately after the first bifurcation point in (a). }\label{fig2}
\end{figure}

Shown in Fig.~\ref{fig2}(a) is the bifurcation diagram of the solution corresponding to
Fig.~\ref{fig1}(a) as a function of the dissipation parameter $\gamma$ for a fixed value of $\Delta$
and $\lambda$. As $\gamma$ increases from zero, there is a saddle-node bifurcation. Following the
existence branches further, we obtain the full diagram as presented in the figure, where there is a
pair of saddle-node bifurcations and the diagram is symmetric with respect to the vertical line
$\gamma=0$. Comparing with the analytical approximation (\ref{rm1}) (for the sake of simplicity we
use (\ref{rm2}) instead) shown as vertical dashed line in the figure, we obtain a satisfactory
agreement. As for the solution profile along the bifurcation diagram, by visual observation only, we
obtain that the solution is shifted to the left in $z$ as $\gamma$ changes and due to the periodicity
returns to its initial position after completing one full bifurcation loop.

In addition to the existence, we also investigated the stability. Despite the symmetry of the
existence diagram, the stability is expectedly asymmetric. The stability of the corresponding
solution is indicated by the solid curve in Fig.~\ref{fig2}(a). In this case, we obtain that the
damping can indeed stabilize solutions. However, this is not necessarily always the case.
When, e.g., the coupling parameter $\lambda$ is large enough, we obtain that all corresponding
solutions along the bifurcation diagram can be unstable. In Fig.~\ref{fig2}(b), we show a typical
dynamics of unstable solutions.

The existence and stability dynamics of solutions due to the perturbation above are rather general
and independent of the period $\overline{T}$. Nevertheless, we observe that the larger the period,
the more unstable the solution. In the same effect as the case of large enough $\lambda$ that we
mentioned above, the presence of dissipation may not be able to stabilize the corresponding solutions
in their entire existence domain. Therefore, using the same argument, 
we expect that the localized wave is always unstable. This is due to the fact that
the solution background, when unperturbed, is a saddle-point. Moreover, the instability in this case
is typically in the form of an unbounded blow-up, similarly to that found in Ref.~\cite{1}.

\subsection{Periodic waves due to drives}

In addition to the periodic and localized waves discussed in the previous subsection,
one also has periodic solutions due to the travelling drive. In this case, the amplitude of the
solutions is proportional to the drive amplitude. Such solutions were not
discussed analytically here.
However, the regularity and uniqueness result of \cite{1} for this type of solutions is immediately
applicable to our case. Therefore, for the sake of completeness, we also
discuss them here.

\begin{figure}[tbhp!]
\begin{center}
\subfigure[]{\includegraphics[width=0.4\textwidth]{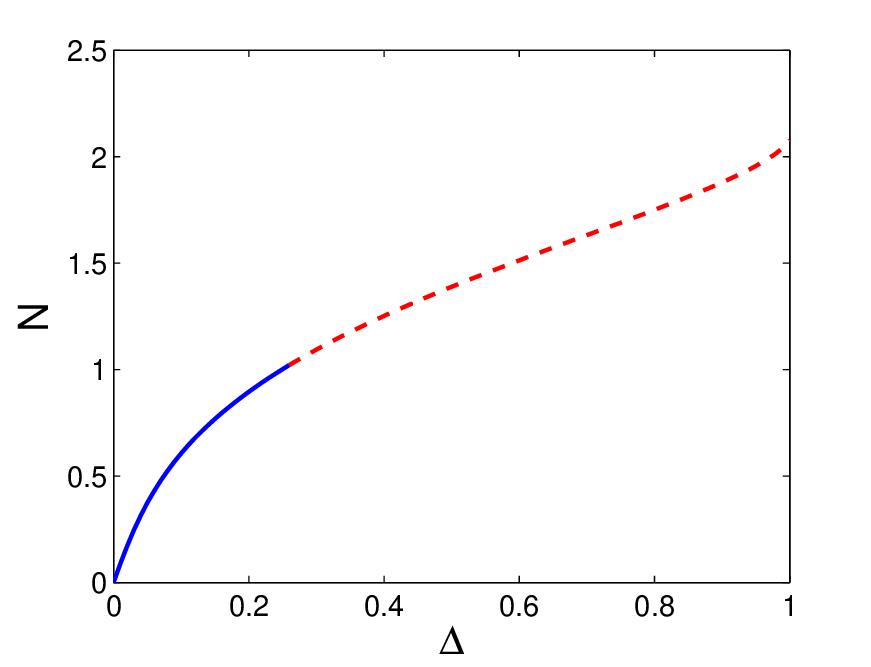}}
\subfigure[]{\includegraphics[width=0.4\textwidth]{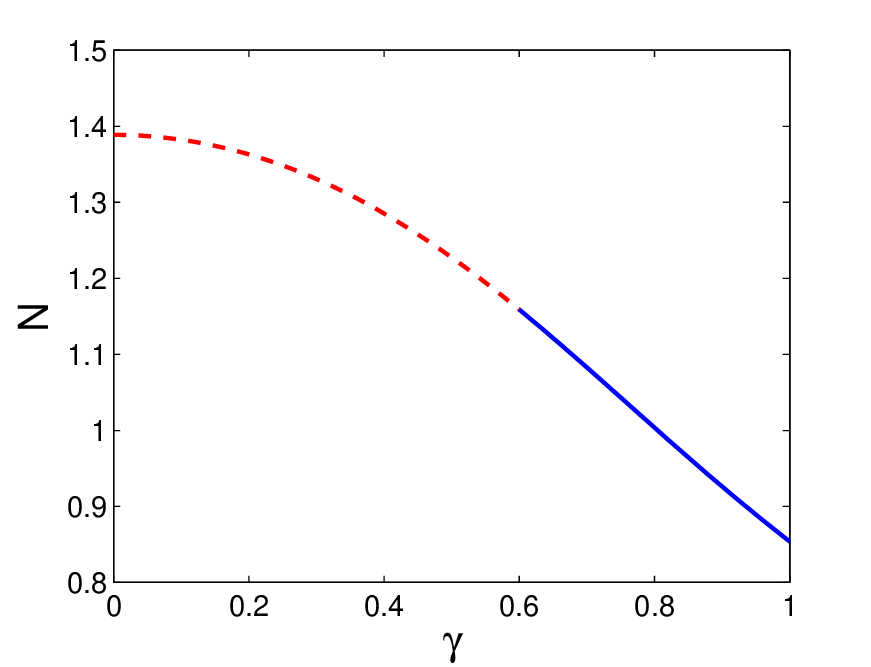}}\\
\subfigure[]{\includegraphics[width=0.4\textwidth]{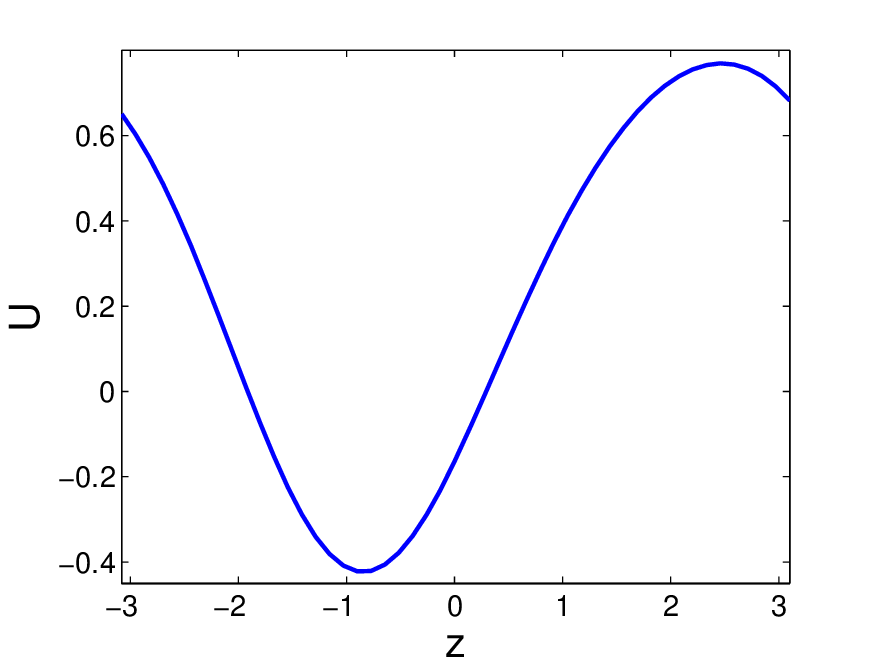}}
\subfigure[]{\includegraphics[width=0.4\textwidth]{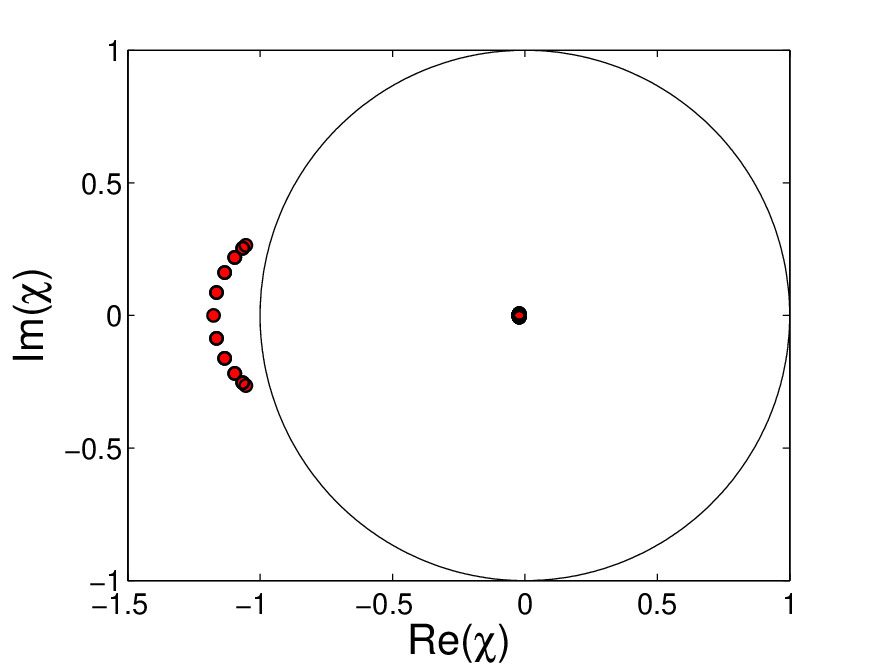}}
\end{center}
\caption{(a-b) The existence curve of periodic waves due to the traveling drive for (a) varying
$\Delta$ with $\gamma=0$ and $\lambda=0.1$; (b) varying $\gamma$ with $\Delta=0.5$ and $\lambda=0.1$.
The vertical axis is the solution norm $N=\sqrt{\int_0^{\overline{T}}U^2(z)\,dz}$. Solid and dashed
curve indicate unstable and stable solutions, respectively. The stability is calculated using $20$
sites. (c) The solution profile with $\Delta=0.5,\,\lambda=0.1,$ and $\gamma=0.57$. (d) The Floquet
multipliers of the solution in panel (c), showing the instability of the solution.}
\label{fig3}
\end{figure}

In Fig.~\ref{fig3}, we show one example of the existence and stability of periodic waves due to the
travelling drive. In panel (a), one can observe that for small enough driving amplitude the solution
is stable. When it is increased further, there is a critical drive amplitude above which the
solution becomes unstable. The reason is the same as that in Fig.~\ref{fig1}, i.e., the drive
creates an instability caused by multipliers leaving the unit circle from $-1$. Introducing the
dissipation $\gamma$, one can stabilize unstable solutions. Shown in panel (b) is the existence and
stability diagram for varying $\gamma$, where the stabilization effect can readily be observed.
In panel (c), we plot the profile of an unstable solution for one set of parameter values. The
multiplier structure in the complex plane is depicted in panel (d).

\begin{figure}[tbhp!]
\begin{center}
{\includegraphics[width=0.5\textwidth]{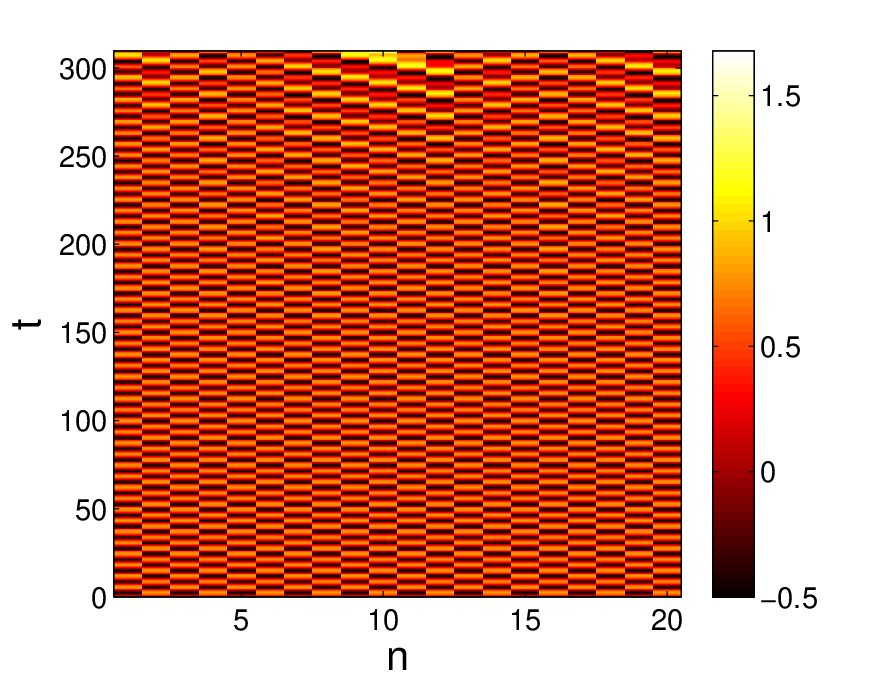}}
\end{center}
\caption{Typical dynamics of the instability of a periodic wave due to the drive. The initial condition corresponds to the solution plotted in Fig.~\ref{fig3}(c).}\label{fig4}
\end{figure}

The time evolution of the unstable solution in Fig.~\ref{fig3}(c) is shown in Fig.~\ref{fig4}.
The dynamics, where one or more sites blow up, 
is typical for this type of waves.

\section{Conclusion}

Using a Melnikov-type analysis we have discussed analytically the existence of periodic and
localized travelling waves in an array of magnetic metamaterials. Intrinsic
characteristics of the system, such as the coupling coefficient and the loss coefficient, affect the
stability of the system. Our analysis showed that even if we have instability, under certain
conditions, we can make our solution stable by adjusting the amplitude of driving. We have also
used direct numerical simulations to compared and confirm the above analytical result.
Moreover, by computing the Floquet multipliers of the periodic solutions we also determined the stability of the
solutions. It was shown that the travelling drive can act both as a stabilizer and as a destabilizer.
In addition, periodic waves with amplitude being proportional to the drive strength were also studied numerically.
In this case, we observed that due to the loss, an unstable solution can be stabilized. 


It would be particularly interesting to follow the lines of the present study and analyze
similar models which appear in the context of nonlinear metamaterials. Indeed, such nonlinear lattice models
appear in the context of nonlinear left-handed transmission line metamaterials \cite{review1} and
have been studied also in experiments (see, e.g., Refs.~\cite{exptl1,exptl2,exptl3}). In most of the relevant
settings, the focus was on the reduction of the lattice model to an effective nonlinear Schr{\"o}dinger (NLS)
equation, by means of which approximate soliton solutions of the original model were presented.
Nevertheless, our analysis may be extremely useful in identifying travelling periodic wave or other
(than envelope solitons) localized solutions, and studying their persistence and stability. Work is in progress
towards this direction and relevant results will be reported elsewhere.

\section*{Acknowledgments}
The work of M.A has been co-financed from resources of the operational program 'Education and Lifelong Learning' of the European Social Fund and the National Strategic Reference Framework (NSRF) 2007-2013 within the framework of the Action <State Scholarships Foundation's  (IKY) mobility grants programme for the short term training in recognized scientific/research centres abroad for candidate doctoral or postdoctoral researchers in Greek universities or research centres>.
The work
of V.R. has been co-financed by the European Union (European Social Fund – ESF) and Greek
national funds through the Operational Program “Education and Lifelong Learning” of the National
Strategic Reference Framework (NSRF) – Research Funding Program: THALES – Investing in
knowledge society through the European Social Fund.
The work of D.J.F. was partially supported by the Special Account for Research Grants of the University of Athens.

\appendix{Elliptic functions}

For the calculation of the Melnikov function we have used the following integral,
involving the square root of three linear factors $\sqrt{t-\alpha} ,\sqrt{t-b}$ and $\sqrt{t-c}$.
One of the limits of the integration will usually be taken as a zero of the polynomial under the radical sign,
while the other limit is considered variable. However, it may easily be used when neither limit is fixed.
$$\int_{a} ^{y} \frac{dt}{\sqrt{(t-\alpha)[(t-b_{1})^{2} +a_{1} ^{2}]}}=g\int_{0} ^{u_{1}} du =g u_{1}=g\cn^{-1} (\cos \phi ,k)$$
where
$$\cn u=\frac{A+\alpha-t}{A-\alpha+t}, \quad k^{2} =\frac{A+b_{1} -\alpha}{2A}, \quad g=\frac{1}{\sqrt{A}},$$
$$A^{2}=(b_{1} -\alpha)^{2} +a_{1} ^{2}, \quad (t-b)(t-c)=(t-b)(t-\overline{b})=(t-b_{1})^{2} +a_{1} ^{2},$$
$$a_{1} ^{2} =-\frac{(b-\overline{b})^{2}}{4}, \quad b_{1}=\frac{b+c}{2} =\frac{b+\overline{b}}{2},$$
$$\phi={\rm am} u_{1} =\cos^{-1} \left(\frac{A+\alpha-y}{A-\alpha+y}\right), \quad \cn u_{1}=\cos \phi.$$
Here, $\alpha$ is real, $b,c$ are complex and $y>\alpha$.

\appendix{The second term of the Melnikov function}

The second term of the Melnikov function is found as follows:

$$\frac{2\pi u}{\overline{T}(\alpha)}\int_0^{\overline{T}(\alpha)} h^{'}\left(\frac{2\pi u}{\overline{T(\alpha)}}z+ a\right)U_{0,\alpha} (z)\,dz$$

$$=-\frac{2\pi u}{\overline{T} (\alpha)}\int_0^{\overline{T} (\alpha)} \sin \left(\frac{2\pi u}{\overline{T} (\alpha)} z+a\right) \left(\frac{A(1-cn(\frac{z}{g}))}{1+cn(\frac{z}{g})} +\alpha\right) \,dz$$

$$=-\frac{2A\pi u}{\overline{T} (\alpha)}\int_0^{\overline{T} (\alpha)} \sin \left(\frac{2\pi u}
{\overline{T} (\alpha)} z+a\right) \left(\frac{(1-cn(\frac{z}{g}))}{1+cn(\frac{z}{g})} \right) \,dz$$

$$ -\frac{2\alpha\pi u}{\overline{T} (\alpha)}\int_0^{\overline{T} (\alpha)} \sin \left(\frac{2\pi u}{\overline{T} (\alpha)} z+a\right) \,dz$$

$$=-\frac{A\pi u}{2K(k)}\int_0^{4K(k)} \sin \left(\frac{2\pi u}
{\overline{T} (\alpha)} z+a\right) \left(\frac{1-cn(z)}{1+cn(z)} \right) \,dz$$

$$=-\frac{A\pi u}{2K(k)}(\sin a \int_{-2K(k)}^{2K(k)} \cos (\frac{\pi u}{2K(k)} z) (\frac{1-cn(z)}{1+cn(z)}) \,dz$$

$$+\cos a \int_{-2K(k)}^{2K(k)}\sin (\frac{\pi u}{2K(k)} z) (\frac{1-cn(z)}{1+cn(z)}) \,dz )$$

$$=-\frac{A\pi u}{2K(k)} \sin a \int_{0}^{4K(k)} \cos (\frac{\pi u}{2K(k)} z) (\frac{1-cn(z)}{1+cn(z)}) \,dz $$

$$=-\frac{A\pi u}{2K(k)} \sin a \int_{0}^{4K(k)} \cos (\frac{\pi u}{2K(k)} z) (-1+\frac{2}{(1+cn(z))})\,dz$$

$$=\frac{A\pi u}{2K(k)} \sin a \Xi.$$

\end{document}